# Disorder, pseudospins, and backscattering in carbon nanotubes

Paul L. McEuen, Marc Bockrath, David H. Cobden*, Young-Gui Yoon, and Steven G. Louie

*Department of Physics, University of California and Materials Sciences Division,
Lawrence Berkeley National Laboratory, Berkeley, California, 94720*

(June 2, 1999)

We address the effects of disorder on the conducting properties of metal and semiconducting carbon nanotubes. Experimentally, the mean free path is found to be much larger in metallic tubes than in doped semiconducting tubes. We show that this result can be understood theoretically if the disorder potential is long-ranged. The effects of a pseudospin index that describes the internal sublattice structure of the states lead to a suppression of scattering in metallic tubes, but not in semiconducting tubes. This conclusion is supported by tight-binding calculations.

Single-wall carbon nanotubes (SWNTs) are two-dimensional (2D) graphene sheets rolled into nanometer-diameter cylinders[1,2] that can either be 1D metals or semiconductors, depending on how the sheet is rolled up. This surprising behavior follows from the unusual band structure of a graphene sheet. It is a semimetal with a vanishing gap at the corners of the first Brillouin Zone (BZ) where the π (bonding) and π* (antibonding) bands touch at two inequivalent wavevectors ***K*** and ***K'*** (Fig. 1(a)). As the Fermi level moves due to chemical or electrostatic doping, the Fermi surface becomes circular arcs at the corners of the BZ, as is shown in Fig. 1(a) for hole doping. This Fermi surface can be more simply represented in the extended zone scheme by piecing together the arcs to form Fermi circles of radius $k$ centered around the ***K*** (***K'***) point. When a graphene sheet is rolled up into a tube, the allowed wavevector components perpendicular to the tube axis become quantized, resulting in 1D subbands with allowed $k$'s represented by dashed lines in Figures 1(b) and (c). For metallic tubes (Fig. 1(b)), one set of allowed wavevectors goes through the ***K*** point and there are propagating modes at $E_f$ at $+k$ and $-k$. This 1D mode has a linear (massless) dispersion, as is indicated in the Figure. For semiconducting tubes (Fig. 1(c)), the allowed wavevectors do not go through the ***K*** point. For small ***k***, there are thus no allowed states at $E_f$, but if the tube is doped sufficiently, the Fermi circle reaches the nearest 1D subband and propagating modes exist, whose (massive) dispersion is shown in the Figure.

A wealth of scanned probe and electrical transport measurements have been performed to probe the electronic structure and conducting properties of SWNTs[2]. Overall, the experimental results agree with the predictions of band structure given above. Many interesting open issues remain, however, particularly concerning the effect Coulomb interactions[3-6] and

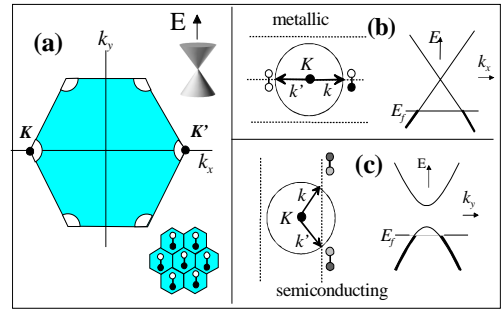

**Figure 1** (a) Filled states (shaded) in the first Brillouin Zone of a single p-type graphene sheet. The sheet contains of two carbon atoms per unit cell (lower right inset). The dispersions of the states in the vicinity of $E_f$ are cones (upper right inset) whose vertices are located at the ***K*** and ***K'*** points. The Fermi circle around the ***K*** point, the allowed $k$ vectors, and their dispersion are shown in (b) and (c) for a metallic and semiconducting tube, respectively. The dumbbells represent the molecular orbitals comprising the states, with white-white, white-black, and gray dumbells representing a bonding, antibonding, and mixed orbitals, respectively.

disorder[7,8,9] on the electronic states. For example, recent theoretical work has emphasized that the effects of disorder may be significantly reduced in SWNTs for a number of reasons[7-9]. Experiments indeed give compelling evidence that a *metallic* tube can have a very long mean free path $\ell$ - on the order of microns[10-14]. Initial experiments on doped *semiconducting* tubes, however, have yielded $\ell$'s that are orders of magnitude shorter[15,16]. This is perhaps surprising, since the tubes are nearly structurally identical and the amount of disorder likely very similar. In this letter, we address this apparent discrepancy between the properties of metallic and doped semiconducting nanotubes.

We begin by discussing the experimental evidence that $\ell$ can be very long in metallic SWNTs. Figure 2 shows a measurement of a nanotube rope ~ 8μm in length. At low $T$, Coulomb oscillations in the conduc-tance $G$ vs. gate voltage $V_g$ are observed as electrons are added to the rope [10, 11]. Using the charging energy $U \sim 0.5$ meV determined from the $T$-dependence, the effective length $L_{eff}$ of the segment of tube to which the electrons are added can be estimated [10, 11]. For this device, we find $L_{eff} \sim 10$ µm, which is approximately the physical tube length, as previously observed by the DELFT group[10]. Note that any significant backscattering within the tube would localize the electronic states on the scale of $\ell$ and effectively break the tube into a series of dots[17]. This would result in multiple Coulomb blockade periods as a function of $V_g$ with larger charging energies. The observation of a single, well-defined, and small charging energy is thus very strong evidence that $\ell$ is many µms in length.

Additional evidence for large $\ell$'s comes from measurements of the two terminal conductance of nanotubes with near-ohmic contacts. For perfect contacts, the conductance is predicted to be: $G = (e^2/h)\sum T_i$, where $T_i$ is the transmission coefficient for each of the four 1D channels propagating through the tube. Measurements by a number of groups [11, 13, 14] have yielded conductances $\sim e^2/h$, indicating that the $T_i$'s can be on the order of unity, even for tubes many microns in length. Clearly, then, metallic tubes can have mean free paths at the micron length scale.

We now turn to experiments on semiconducting tubes. Tans et al. [15] and Martel et al. [16] measured electrostatically doped $p$-type tubes and Bockrath et al.[18] measured $n$-type tubes that were chemically doped. These results can be analyzed using a model of a diffusive conductor. In the simplest version, transport though the tube is limited by scatterers spaced at a distance $\ell$, each with transmission probability $T_i \sim ½$. The conductance of a tube of length $L$ is then: $G \cong (4e^2/h)(\ell/L)$. Using the physical length of the tube and the maximum measured conductance, these experiments indicate $\ell \sim 2$nm at the largest carrier densities. This is *three orders of magnitude* shorter than the $\ell$ found above for metallic tubes.

To investigate this striking discrepancy further, we have performed extensive measurements on semiconducting tubes at both room and low temperatures. Fig. 3 shows the $G$ vs. $V_g$ measured one device. At room temperature, the conductance increases as $V_g$ is decreased and holes are added to the valence band of the semiconducting tube. (The saturation of $G$ at large negative $V_g$ is believed to be due to the contact resistance for tunneling into to the tubes[15, 16].) As $T$ is lowered, $G$ is suppressed and breaks up into a series of peaks as a function of $V_g$. At low temperatures ($T < 20$ K), $G$ is immeasurably small at all $V_g$. The lower inset to Fig. 3 shows the differential conductance, $dI/dV$, for a different semiconducting tube device as a function of $V_g$ and $V$ at $T = 4.2$ K. The data is plotted as a gray scale. There is a gap around the origin where $dI/dV = 0$. This gap shows complex behavior as a function of $V_g$ and is followed by a finite conductance region above $V \sim 25 - 50$ mV. Qualitatively similar results have been obtained on a number of devices consisting of both ropes and single tubes (as determined by AFM measurements of the rope/tube height).

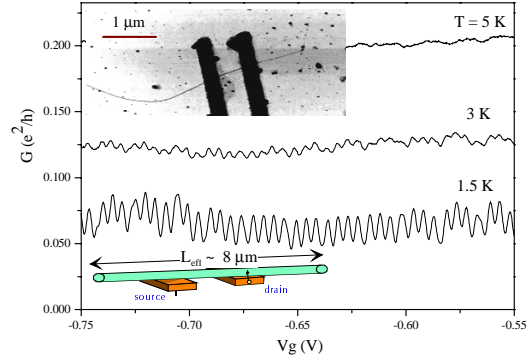

**Figure 2**. Conductance versus gate voltage at different temperatures for the metallic nanotube device shown in the upper inset. The 3 nm diameter and 8 µm long nanotube rope is draped over 2 contacts that make tunnel contact to a metallic tube in the rope. A voltage applied to the doped substrate is used to adjust the carrier density. The appearance of the CB oscillations only at very low temperatures ( ~1.5 K) indicates that the electrons are delocalized over the entire length of the tube, an indicated in the lower inset.

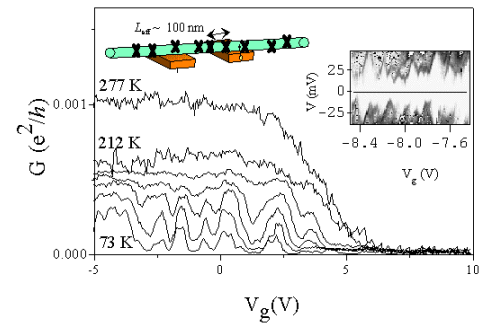

**Figure 3**. $G$ vs. $V_g$ for a semiconducting nanotube device with contacts separated by 0.5 µm. Holes are added to the tube below $V_g = 5$ V and the tube becomes conducting. Irregular Coulomb oscillations are observed below $T \sim 150$ K. The lower inset shows $dI/dV$ vs. $V$ and $V_g$ plotted as a gray scale for a second device at $T = 4.2$ K. Complex structure consistent with transport through a number of quantum dots in series is seen. The T-dependence and typical charging energy indicates that the tube is broken up into segments of length $L_{eff} \sim 100$ nm, as indicated in the schematic.

The data in Fig. 3 are highly reminiscent of measurements of the Coulomb blockade for a number of dots in series[19, 20]. In these systems, an electron must hop through a series of quantum dots, each with a typical charging energy $U$, for current to flow. Since at any $V_g$, some of the dots will be blockaded, $dI/dV = 0$ at low energies. Thermal energies $kT$ or finite bias energies $eV$ on the order $U$ are required to overcome the Coulomb blockade and produce a finite conductance. We therefore conclude that in semiconducting tubes disorder effectively breaks the tube into a series of dots separated by tunnel barriers, as is schematically illustrated in the inset to Fig. 3. The conductance is thus determined by tunneling through a series of quantum dots.

We can estimate the size of these disorder-induced dots from the temperature and bias dependence of the Coulomb blockade features. Since the features appear at energy scales 100 times larger than for the metallic tube in Fig. 2, we find $L_{eff} \sim 100$ nm. Since the device is $\sim 500$ nm long, this implies that the effective sample consists $\sim 5$ dots in series. From the conductance at room temperature, where charging effects are minimal, we estimate that the tunnel barriers between the dots each have transmission probabilites $\sim 0.001$-$0.1$.

These measurements indicate that the diffusive transport model discussed previously – consisting of a large number of scatterers each with $T_i \sim \frac{1}{2}$ - is inappropriate for these samples. Instead, strong disorder over a much longer length scale better describes this system. It is still the case, however, that $G \ll e^2/h$, indicating that semiconducting tubes are much more strongly influenced by disorder than metallic tubes.

To understand this difference, we first review in detail the nature of the electronic states in graphite near $E_f$. The band structure in the vicinity of the $K$ ($K'$) point can be described within the $k \cdot p$ approximation by a 2D Dirac Hamiltonian for massless fermions, $H = \hbar v_F \sigma \cdot k$ [21]. Here $k$ is the wavevector measured relative to the $K$ ($K'$) point and the $\sigma$'s are the Pauli matrices. This Hamiltonian is well-known in both condensed-matter and particle physics; in the latter case, it is used to describe, e.g. a 2D massless neutrino. The states and their corresponding energies are given by[8, 9, 21]:

$$|k> = \frac{1}{\sqrt{2}} e^{ik \cdot r} \begin{pmatrix} -ibe^{-i\theta_k/2} \\ e^{i\theta_k/2} \end{pmatrix}; \quad E = b\hbar v_F |k| \qquad (1)$$

where $\theta_k$ is the angle that $k$ makes with the y-axis in Fig. 1(a) and $b = 1(-1)$ for states above(below) the energy at $K$. Eq. 1 shows that, in addition to their real spin, the electrons possess a pseudospin - a two-component vector that gives the amplitude of the electronic wavefunction on the two sublattice atoms. Inspection of Eq. 1 reveals that the spinor is tied to the $k$ vector such that it always points along $k$. This is completely analogous the physical spin of a massless neutrino which points along the direction of propagation. The states around $K$ correspond to right-handed neutrinos (pseudospin parallel to $k$), whereas those around $K'$ are left-handed (pseudospin antiparallel to $k$). For the antiparticles ($b = -1$) this situation is reversed. Physically, this pseudospin means that the character of the underlying molecular orbital state depends upon the propagation direction. For example, a negative energy state near $K$ with a positive $k_x$ is built from a anti-bonding molecular orbitals while the state with -$k_x$ is built from bonding orbitals. This is schematically indicated in Figure 1(b).

Following Ando and collaborators[8, 9], we now consider scattering between these allowed states in a carbon nanotube due to *long-range disorder*, i.e. disorder with Fourier components $V(q)$ such that $q \ll K$. In this case, the disorder does not couple to the pseudospin portion of the wavefunction since the disorder potential is approximately constant on the scale of the inter-atomic distance. The resulting matrix element between states is then[8]: $|<k'|V(r)|k>|^2 = |V(k-k')|^2 \cos^2(\frac{1}{2}\theta_{k,k'})$, where $\theta_{k,k'}$ is the angle between the initial and final states. The first term is just the Fourier component at the difference in the $k$ values of the initial and final envelope wavefunctions. The *cos* term is the overlap of the initial and final spinor states.

For a metallic tube (Fig. 1(b)), backscattering in the massless subband corresponds to scattering between $|k_x>$ and $|-k_x>$. Such scattering is forbidden, however, since the molecular orbitals of these two states are orthogonal, as was clearly emphasized by Ando et al.[8, 9]. In semiconducting tubes, however, the situation is quite different (Fig. 1(c)). The angle between the initial and final states is $< \pi$, and scattering is thus only partially suppressed by the spinor overlap. As a result, semiconducting tubes should be sensitive to long-range disorder, while metallic tubes should not. Note that *short-range disorder*, $q \sim K$, will couple the molecular orbitals together and lead to scattering in all of the subbands.

To support this picture, we have performed tight-binding calculations of the conductance G of metal and semiconducting tubes in the presence of a scattering potential. We employ the Landauer formalism to calculate the conductance from the transmission coefficients $T_i$ of each subband. A Gaussian disorder potential of the form $V(r) = V_o \exp(-r^2/2\sigma^2)$ centered on one of the atoms on the nanotube wall is included in the tight-binding Hamiltonian. The transmission coefficients are obtained from boundary condition matching between the disorder-free region and the disordered region.

In Figure 4, the calculated $G(E)$ is shown for two realizations of a single Gaussian scatterer with the same integrated strength but different widths corresponding to long-range (dashed lines) and short-range disorder (dash-dot lines). The massless 1D band of a metallic tube is unaffected by a long-range scatter, but there is significant backscattering of the states in the semiconducting tube in the region near the threshold for transmission. There is also backscattering of the higher subband states of the metallic tube, as is expected from extending the arguments above. This calculation clearly demonstrates that the two types of subbands (massive and massless) are affected very differently by long-range disorder in a manner accurately captured by the physics of the pseudospin discussed above.

These theoretical considerations agree very well with the experimental results. Long-length scale disorder due to, e.g. localized charges near the tube, breaks the semiconducting tube into a series of quantum dots with large barriers and a dramatically reduced conductance. Metallic tubes, on the other hand, are insensitive to this disorder and remain near-perfect 1D conductors. In the future, it will be great interest to explore other experimental manifestations of this pseudospin degree of freedom in graphene materials.

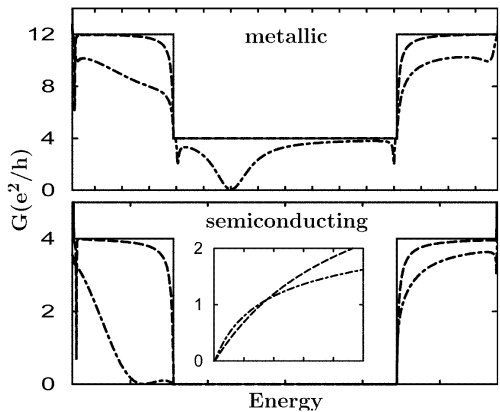

**Figure 4**. Tight-binding calculation of the conductance of an (a) metallic (10,10) tube and (b) semiconducting (17,0) tube in the presence of a Gaussian scatterer. The energy scale on the abscissa is 0.2 eV per division in both graphs. The solid lines show the results for a disorder free tube, while the dash and the dot-dash lines are for, respectively, a single long-range ($\sigma$ = 0.348 nm, $\Delta V$ = 0.5 eV) and short range ($\sigma$ = 0.116 nm, $\Delta V$ = 10 eV) scatterer centered on the wall of the tube. Here $\Delta V$ is the shift in the on-site energy at the potential center. The massless band of the metallic tube is unaffected by the long-range scatterer, unlike the massive bands of the metallic and semiconducting tube. All subbands are influenced by the short-range scatterer. The inset shows an expanded view of the onset of conduction in the semiconducting tube at positive $E$, with each division corresponding to 1 meV. To compare to the experimental data, we estimate that a gate voltage change $V_g$ of 1 V in Figure 3 corresponds to a chemical potential change $E$ of the on the order of 1 meV.

We wish acknowledge useful and enlightening discussions with Dung-Hai Lee and thank the Smalley group for the nanotube material used in these studies. Y. Yoon acknowledges discussions with Hyoung Joon Choi and the use of his tight binding code. This work was supported by DOE, Basic Energy Sciences, Materials Sciences Division, the sp$^2$ Materials Initiative and by NSF DMR-9520554.

\* Present Address: Oersted Laboratory, Neils Bohr Institute, Universitetsparken 5, DK-2100 Copenhagen, Denmark.

[1] M. S. Dresselhaus, G. Dresselhaus, and P. C. Eklund, *Science of Fullerines and Carbon Nanotubes* (Academic, San Diego, 1996).

[2] For a review, see C. Dekker, Physics Today **52**, 22 (1999).

[3] C. Kane, L. Balents, and M. P. A. Fisher, Phys. Rev. Lett. **79**, 5086 (1997).

[4] Y. A. Krotov, D.-H. Lee, and S. G. Louie, Phys. Rev. Lett. **78**, 4245 (1997).

[5] R. Egger and A. O. Gogolin, Phys. Rev. Lett. **79**, 5082 (1997).

[6] M. Bockrath *et al.*, Nature **397**, 598 (1999).

[7] C. T. White and T. N. Todorov, Nature **393**, 240 (1998).

[8] T. Ando, T. Nakanishi, and R. Saito, J. Phys. Soc. Jpn. **67**, 2857 (1998).

[9] T. Ando and T. Nakanishi, J. Phys. Soc. Jpn. **67**, 1704 (1998).

[10] S. J. Tans *et al.*, Nature **386**, 474 (1997).

[11] M. Bockrath *et al.*, Science **275**, 1922 (1997).

[12] S. Frank *et al.*, Science **280**, 1744 (1998).

[13] A. Y. Kasumov *et al.*, Science **284**, 1508 (1999).

[14] H. T. Soh *et al.*, preprint (1999).

[15] S. J. Tans, R. M. Verschueren, and C. Dekker, Nature **393**, 49 (1998).

[16] R. Martel *et al.*, Appl. Phys. Lett. **73**, 2447 (1998).

[17] A. Bezryadin *et al.*, Phys. Rev. Lett. **80**, 4036 (1998).

[18] M. Bockrath *et al.*, preprint (1999).

[19] A. A. M. Staring, H. van Houten, and C. W. J. Beenakker, Phys. Rev. B **45**, 9222 (1992).

[20] V. Chandrasekhar, Z. Ovadyahu, and R. A. Webb, Phys. Rev. Lett. **67**, 2862 (1991).

[21] C. L. Kane and E. J. Mele, Phys. Rev. Lett. **78**, 1932 (1997).